\documentclass[10pt, conference, a4paper]{IEEEtran}

\usepackage[footskip=1cm, left=1.5cm, right=1.5cm, top=1.8cm, bottom=2.5cm]{geometry} 
\usepackage{times,url,comment,cite}
\usepackage{amsmath,amssymb,amsfonts,bbm}
\usepackage{amsthm}
\usepackage{graphicx}
\usepackage{color}
\usepackage{xcolor}
\usepackage{bm}
\usepackage{cleveref}
\usepackage{xspace}
\usepackage{multirow}
\usepackage{tabularx}
\usepackage{caption}
\usepackage{subcaption}
\usepackage{framed}
\usepackage{enumitem}
\usepackage{float}
\usepackage{algorithm}
\usepackage[noend]{algpseudocode}
\usepackage[printonlyused]{acronym}
\usepackage[font=small]{caption}

\colorlet{blue}{black}

\newtheorem{theorem}{Theorem}

\newtheorem{lemma}{Lemma}
\newtheorem{assumption}{Assumption}
\newtheorem{definition}{Definition}
\allowdisplaybreaks

\makeatletter

\algnewcommand{\IfThenElse}[3]{
  \State \algorithmicif\ #1\ \algorithmicthen\ #2\ \algorithmicelse\ #3}

\algnewcommand{\IfThen}[2]{
  \State \algorithmicif\ #1\ \algorithmicthen\ #2}
  
\algnewcommand{\ForDo}[2]{
  \State \algorithmicfor\ #1\ \algorithmicdo\ #2
}

\def\argmax{\operatornamewithlimits{arg\!\,max}}
	
\begin{document}
\title{
Constrained Network Slicing Games: \\ Achieving service guarantees and network efficiency 
}
\author{\IEEEauthorblockN{
Jiaxiao Zheng\IEEEauthorrefmark{1},
Gustavo de Veciana\IEEEauthorrefmark{1},
Albert Banchs\IEEEauthorrefmark{2}
}
\IEEEauthorblockA{\IEEEauthorrefmark{1}The University of Texas at Austin, TX}
\IEEEauthorblockA{\IEEEauthorrefmark{2}University Carlos III of Madrid \& IMDEA Networks Institute, Spain}
}

\newif\ifappendix
\appendixfalse
\appendixtrue

\newif\ifextended
\extendedtrue

\maketitle
\thispagestyle{plain}
\pagestyle{plain}

\begin{abstract}
Network slicing is a key capability for next generation mobile networks. It enables one to cost effectively customize logical networks over a shared infrastructure. A critical component of network slicing is resource allocation, which needs to ensure that slices receive the resources needed to support their mobiles/services while optimizing network efficiency. In this paper, we propose a novel approach to slice-based resource allocation named \emph{Guaranteed seRvice Efficient nETwork slicing} (GREET). The underlying concept is to set up a {\em constrained} resource allocation game, where ($i$) slices unilaterally optimize their allocations to best meet their (dynamic) customer loads, while ($ii$) constraints are imposed to guarantee that, if they wish so, slices receive a pre-agreed share of the network resources. The resulting game is a variation of the well-known Fisher market, where slices are provided a budget to contend for network resources (as in a traditional Fisher market), but (unlike a Fisher market) prices are constrained for some resources to provide the desired guarantees. In this way, GREET combines the advantages of a share-based approach (high efficiency by flexible 
sharing) and reservation-based ones (which provide guarantees by assigning a fixed amount of resources). 
We characterize the Nash equilibrium, best response dynamics, and propose a practical slice strategy with provable convergence properties. Extensive simulations exhibit substantial improvements over network slicing state-of-the-art benchmarks.

\end{abstract}

\section{Introduction}

There is consensus among the relevant industry and standardization communities.
that a key element in 5G 
mobile networks is \emph{network slicing}. This technology allows the network infrastructure to be ``sliced'' into logical networks, which are operated by different entities and may be tailored to support specific mobile services. This provides a basis for efficient infrastructure sharing among diverse entities, such as mobile network operators relying on a common infrastructure managed by an infrastructure provider, or new players that use a {\em network slice} to run their business (e.g., an automobile manufacturer providing advanced vehicular services, or a city hall providing smart city services). In the literature, the term \emph{tenant} is often used to refer to the owner of a network slice.

A network slice is a collection of resources and functions that are orchestrated to support a specific service. This includes software modules running at different locations as well as the nodes' computational resources, and communication resources in the backhaul and radio network. By tailoring the orchestration of resources and functions of each slice according to the slice's needs, network slicing enables tenants to share the same physical infrastructure while customizing the network operation according to their market segment's characteristics and requirements. 



One of the key components underlying network slicing is the underlying framework for 
\emph{resource allocation}: we need to decide how to assign the underlying infrastructure resources to each slice at each point in time. When taking such decisions, two major objectives are pursued: ($i$) meeting the customers' needs specified by slice-based Service Level Agreements (SLAs), and ($ii$)  realizing efficient infrastructure sharing by maximizing the overall level of satisfaction across all slices.
Recently, several efforts have been devoted to this problem. 
Two different types of approaches have emerged in the literature: \\
\noindent
\underline{\emph{Reservation-based schemes}}~\cite{aanetworkslicing,globe17,Sci17,huawei,deepcog,Beg17,tmc,orion} where a tenant issues a reservation request 
with a certain periodicity or on demand. Each request involves a given allocation for each resource in the network (where a resource can be a base station, a cloud server or a transmission link). \\ 
\noindent
\underline{\emph{Share-based schemes}}~\cite{ZCVB17,ZCD18,GPSB15,infocom17,ton,twcg,oroton}
where a tenant does not issue reservation requests for individual resources, but rather purchases a share of the whole network. This share is then mapped dynamically to different allocations of individual resources depending on the tenants' needs at each point in time.


\textcolor{blue}{
These approaches have advantages and disadvantages. 
Reservation-based schemes are in principle able to guarantee that a slice's 
requirements are met, but to be efficient, require constant updating of the resource allocations 
to track changing user loads, capacities and/or demands. The overheads of doing so 
at a fine granularity can be substantial: including challenges with maintaining 
state consistency to enable admission control, modifying reservations 
and addressing handoffs. Indeed these overheads are already deemed 
high for basic horizontal and/or vertical handoffs.  As a result, resource allocations need 
to be done at a coarser granularity and slower time-scales resulting in reduced overall efficiency/performance.
}


\textcolor{blue}{
In contrast to the above, in share-based approaches a slice is given a coarse grain share of the network
resources which combined with a fine grain dynamic policy can track rapid changes in a slices' load distributions. 
Indeed, as these schemes do not involve explicit per resource reservation requests, they can more 
rapidly adapt allocations to the demand variations of network slices 
(see, e.g., \cite{son2}). Their main drawback, however, is that tenants 
do not have a guaranteed allocation at individual resources, and as a consequence 
one cannot ensure that slices' requirements will always be met.
}



\vspace{0,125 cm}{\em Key contributions}: 
In this paper, we propose a novel approach to resource allocation among network slices 
named \emph{Guaranteed seRvice Efficient nETwork slicing} (GREET). 
GREET combines the advantages of the above two approaches while avoiding
their drawbacks. The key idea is that a slice is guaranteed a
given allocation at each individual resource, as long as the
slice needs such an allocation, while the remaining resources
are flexibly and efficiently
shared. In this way, GREET is able to provide
guarantees and thus meet the SLA requirement of each slice,
and at the same time it provides a flexible sharing of resources
across slices that leads to an overall optimal allocation. Our key contributions are as follows:
\begin{itemize}
\item We propose the {\em GREET slice-based resource allocation framework}, which relies on a {\em constrained} resource allocation game where slices can unilaterally optimize their allocations under some constraints which guarantee that slices are entitled to a pre-agreed share of the individual network resources specified in their SLAs (Section~\ref{sec-model}).
\item We analyze the resulting network slicing game when slices contend for resources to optimize their performance. We show that the game has a Nash Equilibrium (NE) but unfortunately the Best Response Dynamics (BRD) may not converge to this equilibrium (Section~\ref{sec-analysis}).
\item We propose a {\em GREET strategy for individual slices} that complements our resource allocation framework. The proposed strategy is simple and provides a good approximation to the slice's best response. We show conditions for convergence with the proposed strategy (Section~\ref{sec-greet}). 
\item We perform a simulation-based evaluation confirming that GREET combines the best features of reservation-based approaches, providing service guarantees, and share-based ones, maximizing overall performance (Section~\ref{sec-setup}).
\end{itemize}

\ifappendix
\else
{\color{blue}
Due to space constraints, we refer the reader to \cite{extended-this} for the proofs of 
the theoretical results as 
well as for some additional results.
}
\fi

\section{Resource Allocation Approach}\label{sec-model}

In this section we introduce both the system model and the resource allocation framework proposed in this paper.
 
\subsection{System model}

We consider a set of resources $\mathcal{B}$  shared by 
a set of slices $\mathcal{V}$, with cardinalities $B$ and $V$, respectively. 
$\mathcal{B}$ may denote a set of base stations as well as any other sharable resource type, 
e.g., servers providing compute resources.
While our analysis can be applied to different resource types, in what follows we focus on radio resources and refer to $b \in \mathcal{B}$ as a base station.

We assume that each network slice supports a collection of mobile users, possibly with heterogeneous requirements, each of which is associated with a single base station. The overall set of users on the network is denoted by $\mathcal{U}$, those supported by slice $v$ are denoted by $\mathcal{U}^v$, those associated with base station $b$ are denoted by $\mathcal{U}_b$, and we define $\mathcal{U}^v_b := \mathcal{U}_b \cap \mathcal{U}^v$. The set of active slices at base station $b$, corresponding to those that have at least one user at $b$, is denoted by $\mathcal{V}_b$ (i.e., $|\mathcal{U}^v_b| > 0$ holds for $v \in \mathcal{V}_b$). 

The goal in this paper is to develop a mechanism to allocate resources amongst slices. To that end, we let $f_b^v$ denote the fraction of resources at base station $b$ allocated to slice $v$. We adopt a generic formulation based on divisible resources that can be applied to a variety of technologies. The 
specific resource notion will depend on the underlying technology; for instance, in OFDM resources refer to physical resource blocks, in FDM to bandwidth and in TDM to the fraction of time.

The resources of a base station allocated to a slice are subdivided among the slice's users at the base stations, such that a user $u \in \mathcal{U}^v_b$ receives a fraction $f_u$ of the resource, where $\sum_{u \in \mathcal{U}_b^v}f_u = f_b^v$. With such an allocation, user $u$ achieves a service rate $r_u = f_u \cdot c_{u}$, where $c_{u}$ is the user's achievable rate, defined as the rate that the user would see if she had the entire base station provisioned to herself. Note that $c_u$ depends on the modulation and coding scheme selected for the user given the current radio conditions, which accounts for noise as well as the interference from the neighboring base stations. 
Following similar analyses in the literature (see e.g., \cite{Mah13}), we shall assume that $c_u$  is fixed for each user at a given time.


The focus of this paper is on {\em slice-based resource allocation}: our problem is to decide which fraction of the overall resources we allocate to each slice (e.g., the number of resource blocks of each base station). In order to translate slice-based allocations to specific user-level allocations, the system will further need to decide ($i$) which specific resources will be assigned to each slice, and ($ii$) in turn, the assignment of slice resources to active users. This corresponds to a user-level scheduling problem which is not in the scope of this paper, but may impact the users' achievable rates $c_u$ {\color{blue}(this problem has been addressed, for instance, in~\cite{oroinfocom,mandelli,ksentini}).}

In line with standard network slicing frameworks~\cite{Ric16}, the approach studied in this paper can be flexibly 
combined with different algorithms for user-level allocations. 
The specific mechanism to assign resources to slices is the responsibility of the infrastructure provider, 
which may take into account, e.g., the latency requirements of the different slices. 
The sharing of the resources of a slice amongst its users is up to the slice, 
and different slices may run different scheduling algorithms depending on the requirements of their users. 
For instance, slices with throughput-driven services may opt for opportunistic schedulers~\cite{AsM13,borst2009mobility,kushner2004conv} 
while other slices with latency requirements may opt for delay-sensitive schedulers~\cite{kalil2015qos}.

Depending on its type of traffic, a slice may require different allocations. 
For instance, a URLLC slice with high reliability and/or low latency requirements
may require a resource allocation much larger than its average load, to make make sure sufficient resources
are available and/or delays are low.
By contrast, a slice with eMBB traffic may not require guarantees at each individual 
base station, but may only need a certain average fraction of resources over time for its users (i.e., $f_u$). 

\subsection{GREET: Slice-based Resource Allocation}\label{sec:resource-allocation}

Below, we propose a slice-based resource allocation scheme that, on the one hand, ensures that each slice 
is guaranteed, {\em as needed}, a pre-agreed fraction of the resources at each individual base station, and, on the other hand, 
enables slices to contend for spare resources.
Such division into guaranteed resources and extra ones is in line with current cloud models~\cite{Mattess2010,amazon,google}.
In order to regulate the resources to which a network slice is entitled, as well as the competition for the `excess' resources, we rely on the different types of \emph{shares} defined below. Such shares are specified in the slices' SLAs.

{\color{blue}
\begin{definition} For each slice $v$, 
we define the following pre-agreed \emph{static} shares of the network resources.
\begin{enumerate} 
\item  We let the {\bf {\em guaranteed (resource) share}} $s^v_b$ denote the
fraction of $b$'s resources guaranteed to slice $v$, which must satisfy 
$\sum_{v \in \mathcal{V}}{s^v_b} \leq 1$ in order to avoid over-commitment.  
\item We let $e^v$ denote the  {\bf {\em share of excess resources}}
which slice $v$ can use to contend for the spare network resources.
\item We let  $s^v$ denote the slice $v$'s {\bf {\em overall share}}, 
given by $s^v = \sum_{b \in \mathcal{B}}{s^v_b} + e^v$.
\end{enumerate}
\end{definition}

After being provisioned a fraction of network resource, each slice $v$ has the option to divide its 
own share to
its individual users. This can be done by designating a weight $w_u$ for user $u\in\mathcal{U}^v$. We 
let $\mathbf{w}^v = (w_u, u \in {\cal U}^v)$ denote the weight allocation of Slice $v$ such that
$\|\mathbf{w}^v\|_1 \le s^v$. 
The set of feasible weight allocations is given by $\mathcal{W}^v := \{ \mathbf{w}^v :  \mathbf{w}^v  \in  \mathbb{R}_{+}^{|{\cal U}^v|}~\mbox{and}~\sum_{u\in\mathcal{U}^v}w_u \leq s^v\}$. 
Then, we'll have $l^v_b = \sum_{u\in\mathcal{U}^v_b} w_u$ as the slice $v$'s aggregate 
dynamic \emph{local}
bid to BS $b$, which is determined by its user distribution and must satisfy
that $\sum_{b\in\mathcal{B}} l^v_b \le s^v$. 
We further let $l_b := \sum_{v\in\mathcal{V}_b} l_b^v$ denote the overall bid at resource 
$b$ and $l^{-v}_b := \sum_{v' \neq v } l_b^{v'}$ such bid excluding slice $v$. 
Then if we define 
$\Delta^v_b := \left( l^v_b - s^v_b \right)_+$ as the excessive bid per BS of slice $v$, 
our proposed resource allocation mechanism works as follows.

\begin{definition}
({\em {\bf GREET slice-based resource allocation}})
We determine the fraction of 
each resource $b$ allocated to slice $v$, $(f^v_b, v \in {\cal V}, b \in {\cal B})$, as follows. 
If $l_b \le 1$, then 
\begin{equation}
\label{underload-allocation}
f^v_b = \frac{l^v_b}{l_b}, 
\end{equation}
\mbox{and otherwise}
\begin{align}
f^v_b = \begin{cases}
l^v_b, & l^v_b < s^v_b, \label{overload-allocation} \\
s^v_b + 
\frac{\Delta^v_b}{\sum\limits_{v^\prime \in \mathcal{V}_b} \Delta^{v^\prime}_b}
\left( 1 - \sum\limits_{v^\prime \in \mathcal{V}_b} \min\left(s^{v^\prime}_b, l^{v^\prime}_b\right) \right), &  l^v_b \ge s^v_b.
\end{cases}
\end{align}
\end{definition}
}

The rationale underlying the above mechanism is as follows. 
If $l_b \leq 1$, then (\ref{underload-allocation}) ensures that each slice gets a fraction of 
resources $f^v_b$ exceeding its {\color{blue}local bid} $l^v_b$ at resource $b$.
If $l_b > 1$, then (\ref{overload-allocation}) ensures that a slice whose {\color{blue} local bid} at $b$ 
is less than its guaranteed share, i.e., $l^v_b \leq s^v_b$, receives exactly its local bid, 
and a slice with a {\color{blue}local bid} exceeding its guaranteed share, i.e.,  $l^v_b >  s^v_b$, receives
its guaranteed share $s^v_b$ plus a fraction of the {\em extra resources} proportional to the 
{\color{blue} excessive bid $\Delta^v_b$}.
The extra resources here correspond to those not allocated based on guaranteed resource shares. 
As a slice can always choose a local-bid allocation at resource $b$, $l^v_b$, exceeding its guaranteed
 share, $s^v_b$, this ensures that, if it so wishes, a slice can always attain its guaranteed resource shares.

{\color{blue}
The above specifies the slice allocation per resource. Based on the $w_u$'s, the slices then
allocate base stations' resources to users in proportion to their weights, i.e., 
$
f_u = \frac{w_u}{\sum_{u' \in \mathcal{U}_b^v} w_{u'}}f^v_b,
$
where $f_u$ is the fraction of resources of base station $b$ allocated to user $u \in {\cal U}^v_b$.
}

\ifappendix
{\color{blue}
Following is a simple example demonstrating GREET resource allocation. Suppose we have $V=2$ slices,
and $B=2$ BSs. At BS 1, we have $l^1_1 = l^2_1 = 0.5$, together with $s^1_1 = 0.25$, $s^2_1 = 0.75$.
At BS 2, we have $l^1_2 = 0.25$ and $l^2_2 = 1$, when $s^1_2 = s^2_2 = 0.5$. Such share distribution results in
$e^1 = s^1 - (s^1_1 + s^1_2) = 0$ and $e^2 = s^2 - (s^2_1 + s^2_2) = 0.25$.
The resource provisioning, according to GREET, will be as follows. At BS 1, since $l^1_1 + l^2_1 \le 1$,
$f^1_1 = f^2_1 = 0.5$ per Eq. \eqref{underload-allocation}. While at BS 2, Eq. \eqref{overload-allocation}
takes effect because $l^1_2 + l^2_2 > 1$. For Slice 1, $f^1_2 = l^1_2 = 0.25$ because $l^1_2 < s^1_2$.
For Slice 2, $f^2_2 = 0.75$ and it falls into the case that $l^2_2 \ge s^2_2$.
}
\fi

One can think of the above allocation in terms of market pricing schemes as follows. 
The share $s^v$ can be understood the budget of player $v$ and the local bid $l^v_b$ as the 
bid that this player places on resource $b$. Then, the case where $l_b \le 1$ corresponds to the well-known 
Fisher market~\cite{propresp}, where the price of 
the resource is set equal to the aggregate bids from slices, making allocations proportional to the slices'
bids. GREET deviates from this when $l_b \geq 1$ by modifying the `pricing' as follows: 
for the first $s^v_b$ bid of slice $v$ on resource $b$, GREET sets the price to 1, to ensure that the slice 
budget suffices to buy the guaranteed resource shares. Beyond this, the remaining resources 
are priced higher, as driven by the corresponding slices' excess bids. 

In summary, the proposed slice-based resource allocation scheme 
is geared at ensuring a slice will, if it wishes, be able to get its guaranteed resource
shares, $s^v_b$,  but it also gives a slice the flexibility to contend for excess resources,
by shifting portions of its overall share $s^v$ (both from the guaranteed and excess shares) 
across the network resources, to better meet its current users' requirements {\color{blue} by
aligning with its user traffic}. 
Such a slice-based resource sharing model provides the benefit of protection guarantees as well as 
the flexibility to adapt to user demands. 

\section{Network Slicing Game Analysis}\label{sec-analysis}

Under the GREET resource allocation scheme, each slice must choose how to subdivide 
its overall share amongst its users. Then, the network decides how to allocate base 
station resources to slices. 
This can be viewed as a {\em network slicing game} where, depending on the choices 
of the other slices, each slice chooses an allocation of local bid to base stations that maximizes 
its utility. In this section, we study the behavior of this game; 
we first provide a model for the utility of a slice and then analyze the resulting game. 

\subsection{Slice and Network Utilities}

Note that the users' rate allocations, $(r_u : u \in {\cal U})$, can be expressed as a function of the 
overall slice weight assignments across the network, $\mathbf{w} = (w_u : u \in {\cal U})$. Indeed, 
the weights provide the local bid of each slice at each base station, which determine the resources 
of each slice, as well as the division of such resources across the slice's users at the base station. 
Accordingly, in the sequel we focus the game analysis on the weights and express the resulting 
user rates as $r_u(\mathbf{w})$.

We assume that each slice has a {\em private} utility function, denoted by $U^v$, that reflects 
the slice's preferences based on the needs of its users. 
We suppose the slice utility is simply a sum of its users individual utilities, $U_u$, i.e.,
$
U^v(\mathbf{w}) = 
\sum_{u \in \mathcal{U}^v}{U_u(r_u(\mathbf{w}))}.
$

Following standard utility functions~\cite{She95}\cite{HZC07}, we assume that 
for some applications, a user $u\in\mathcal{U}^v$ may require a guaranteed rate $\gamma_u$, 
hereafter referred to as the user's \emph{minimum rate requirement}. We model the utility functions for rates above the minimum requirement as follows:
\begin{equation*}
U_u(r_u(\mathbf{w})) = 
\begin{cases}
\phi_u F_u(r_u(\mathbf{w})-\gamma_u), \ \ r_u(\mathbf{w}) > \gamma_u,\\
~~~~~~-\infty ~~~~~~~~~~~~~~~ \textrm{otherwise},\\
\end{cases}
\end{equation*}
where $F_{u}(\cdot)$ is the utility function associated with the user, 
and $\phi_u$ reflects the {\em relative priority} that slice $v$ 
wishes to give user $u$, with $\phi_u \geq 0$ and $\sum_{u \in \mathcal{U}^v} \phi_u =1$. 

For $F_u(\cdot)$, we consider the following widely accepted family of functions, referred to as $\alpha$-fair utility functions~\cite{Mo00}:
\begin{equation*}\label{eq-alpha}
F_u(x_u)=\begin{cases}
\frac{(x_u)^{1-\alpha^v}}{(1-\alpha^v)},\quad \alpha^v \neq1 \\
\log(x_u),\quad \alpha^v=1,
\end{cases}
\end{equation*}
where the $\alpha^v$ parameter sets the level of concavity of the user utility functions, which in turn determines the underlying resource allocation criterion of the slice. Particularly relevant cases are $\alpha^v = 0$ (maximum sum), $\alpha^v = 1$ (proportional fairness), $\alpha^v = 2$ (minimum potential delay fairness) and $\alpha^v \to \infty$ (max-min fairness). 

Note that the above utility is flexible in that it allows slice utilities to capture users with different types of traffic:
\begin{itemize}
\item \emph{Elastic traffic} ($\gamma_u = 0$ and $\phi_u > 0$): users with no minimum rate requirements and a 
utility that increases with the allocated rate, possibly with different levels of concavity given by $\alpha^v$.
\item \emph{Inelastic traffic} ($\gamma_u > 0$ and $\phi_u = 0$): users that have a minimum 
rate requirement but do not see any utility improvement beyond this rate.
\item \emph{Rate-adaptive traffic} ($\gamma_u > 0$ and $\phi_u > 0$): users with a 
minimum rate requirement which see a utility improvement if they receive an 
additional rate allocation
above the minimum.
\end{itemize}

Following~\cite{ZCVB17,ZCD18,mora,infocom17,ton,twcg}, we define the overall (network) utility as the sum of the individual slice utilities weighted by the respective overall shares,
\begin{equation}\label{eq:def-social-utility}
U(\mathbf{w}) = \sum_{v\in\mathcal{V}} s^v U^v(\mathbf{w}),
\end{equation}
and the social optimal weight allocation $\mathbf{w}^{\textrm{so}}$
as the allocation maximizing the overall utility $U(\mathbf{w})$, i.e., 
\begin{equation}\label{eq:def-social-optimal}
\mathbf{w}^{\textrm{so}} = \argmax_{\mathbf{w}}~U(\mathbf{w}).
\end{equation}

\subsection{Network Slicing Resource Allocation Game}

Next we analyze the network slicing game resulting from 
the GREET resource allocation scheme and the above slice utility. 
We formally define the network slicing game as follows, where $\mathbf{w}^v$ 
denotes slice $v$ users' weights.


\begin{definition} {\bf {\em  (Network slicing game)}} 
Suppose each slice $v$ has access to the guaranteed shares and the 
{\color{blue} local bid} allocations of the other slices, i.e., 
$s^{v'}_b, l^{v'}_b, v'\in {\cal V}\setminus\{v\}, b \in {\cal B}.$
In the network slicing game, slice $v$ chooses its own user weight allocation $\mathbf{w}^v$ in its strategic space $\mathcal{W}^v$ 
so as to maximize its utility, given that the network uses a GREET slice-based resource allocation. This choice is known
as slice $v$'s Best Response (BR). 
\end{definition}

In the sequel we consider scenarios where the guaranteed shares suffice 
to meet the minimal rate requirements of all users.  
The underlying assumption is that a slice would provision a sufficient shares and/or
perform admission control so to limit the number of users.
We state this formally as follows:


\begin{assumption}\label{assm:well-dimension}
{\bf {\em (Well dimensioned shares)}}
We assume that the minimum rate requirements of the users of all slices
can be met with the slices' guaranteed share at each base station. 
In particular, we assume that $\sum_{u\in\mathcal{U}_b^v}{\underline{f}_u} \le s_b^v$ for all $v \in {\cal V}$ and $b \in {\cal B}$, where $\underline{f}_u = \frac{\gamma_u}{c_u}$ is the minimum fraction of resources required by user $u$ to meet the minimum rate requirement $\gamma_u$. 
When this assumption holds, we say that the (guaranteed) shares of all slices are well dimensioned.
\end{assumption}

The following lemma clarifies that, when the above assumption holds, a slice's best response is 
determined as the solution to a
convex problem and meets the minimum rate requirements of all its users. 
Thus, this result guarantees that, as long as the shares of a slice are properly provisioned, the 
proposed scheme meets the slice's requirements.

\begin{lemma}\label{lm-convex}
When Assumption \ref{assm:well-dimension} holds, 
computing the Best Response under GREET-based resource allocation is a convex optimization problem.
Furthermore, the minimum rate requirements of all the slice's users are satisfied by the Best Response.
\end{lemma}






To characterize the system, it is desirable to determine the existence of a NE. The result below shows that, when the slice shares are well dimensioned, if we impose that weights have to be above some value $\delta$ (which can be arbitrarily small), the existence of a NE is guaranteed. However, if we do not impose such lower bound on the weights, a NE may not exist.

\begin{theorem}\label{thm:ne-existence}
Suppose that Assumption~\ref{assm:well-dimension} holds 
and that we  constrain user weights to be positive, i.e., for all $u \in {\cal U}$ 
$w_u \ge \delta$ for some $\delta >0$. Then, a NE exists. 
However, if we do not impose this constraint on the weights, 
an NE may not exist. 
\end{theorem}

Beyond the existence of equilibria, it is also desirable to have a 
dynamic behavior that leads to an equilibrium. 
Below, we analyze the Best Response Dynamics (BRD), 
where slices update their Best Response sequentially, one at a time, in a Round Robin manner. 
Ideally, we would like this process to converge after a sufficiently large number of rounds. 
However, the following result shows that this need not be the case. 

\begin{theorem}\label{thm-conv}
Suppose that Assumption~\ref{assm:well-dimension} holds 
and that we  constrain user weights to be positive, i.e., for all $u \in {\cal U}$ 
$w_u \ge \delta$ for some $\delta >0$.
Then, even though a NE exists, the Best Response Dynamics may not converge.
\end{theorem}

\section{GREET Slice Strategy}\label{sec-greet}

In addition to the equilibrium and convergence issues highlighted in 
Theorems~\ref{thm:ne-existence} and~\ref{thm-conv}, a drawback of the 
Best Response algorithm analyzed in Section~\ref{sec-analysis} is 
its complexity. Indeed, to determine its best response, a slice needs to solve a convex optimization problem. 
This strays from the simple algorithms, both in terms of implementation
and understanding, that get adopted in practice and tenants tend to prefer.
In this section, we propose an alternative slice strategy to the best response, which we refer to as the \emph{GREET share allocation policy}. This policy complements the resource 
allocation mechanism proposed in Section~\ref{sec-model}, leading to the overall GREET framework consisting of two pieces: the resource allocation mechanism and the share allocation policy.

\subsection{Algorithm definition and properties}

The GREET resource allocation given in Section~\ref{sec-model} depends on the bid that slices allocate at each base station. In the following, we propose the \emph{GREET share allocation policy} to 
determine how each slice allocate its share across its users and resources. Our proposal works on the basis of user weights, corresponding to the share fraction allocated to individual users: we first determine the weights of all the users of the slice, and then compute the local bid by summing the weights of all the users at 
each base station, i.e., $l_b^v = \sum_{u \in \mathcal{U}_b^v}{w_u}$.

Under the proposed GREET share allocation, slices decide the weight allocations of their users based on two parameters: one that determines the minimum allocation of a user ($\gamma_u$) and another one that determines how extra resources should be prioritized ($\phi_u$). A slice first assigns each user $u$ the weight needed to meet its minimum rate requirement $\gamma_u$. Then, the slice allocates its remaining  share amongst its users in proportion to their priority $\phi_u$. The algorithm is formally defined below. Note that this algorithm does not require revealing 
each slices' local bids to the others but only aggregates, 
which discloses very limited information about slices' individual sub-shares and leads to low signaling
overheads. 

\begin{definition} {\bf {\em (GREET Share Allocation)}}
Suppose that each slice $v$ has access to the following three aggregate values for each base station: $l_b^{-v}$, 
$\sum_{v' \in \mathcal{V}_b\setminus\{v\}}\Delta^v_b$ and $\sum_{v'\in\mathcal{V}_b\setminus\{v\}} \min( s^{v'}_b, l^{v'}_b)$. Then, the GREET share allocation is given by the weight computation determined by Algorithm~1.  
\end{definition}

\begin{algorithm}[t]
\caption{GREET share allocation round for slice $v$}
\begin{algorithmic}[1]
\ForDo{user $u\in\mathcal{U}^v$}
	  {set $\underline{f}_u \leftarrow  \frac{\gamma_u}{c_u}$}
\ForDo{each base station $b\in\mathcal{B}$}
	  {set $\underline{f}^v_b \leftarrow \sum_{u\in\mathcal{U}^v_b}\underline{f}_u$}
\For{user $u\in\mathcal{U}^v$}
	\If{$l_b^{-v} + \underline{f}^v_b \le 1 $}
		{set $\underline{w}_u \leftarrow \frac{\underline{f}_u}{1 - \underline{f}^v_b} l_b^{-v}$}
	\Else
		\If{$s^{v}_b \ge \underline{f}^v_b$}
		   {set $\underline{w}_u \leftarrow \underline{f}_u$}
		\Else
			 { set $\underline{w}_u \leftarrow$ expression given by \eqref{eq:weight-insufficient-guarantee}}
		\EndIf
	\EndIf
\EndFor
\If{$\sum_{u \in \mathcal{U}^v}{\underline{w}_u} \le s^v$}
	\For{user $u\in\mathcal{U}^v$}
		\State set $w_u \leftarrow \underline{w}_u	+ \phi_u 
                  \left(s^v - \sum_{u' \in \mathcal{U}^v}{\underline{w}_{u'}}\right)$
	\EndFor
\Else
	\While{$\sum_{u \in \mathcal{U}^b}{w_u} \le s^v$}
		\State select users in order of increasing $\underline{w}_u$
		\State set $w_u \leftarrow \underline{w}_u$
	\EndWhile 
\EndIf
\end{algorithmic}
\end{algorithm}
\vspace{-0.25em}

Algorithm~1 realizes the basic insight presented earlier. The slice, say $v$, first computes 
the minimum resource allocation required to satisfy 
the minimum rate requirement of each user, denoted by $\underline{f}_u$. These are then summed to obtain 
the minimum aggregate requirement at each base station, denoted by $\underline{f}^v_b$ (see Lines 1-2 of the algorithm). 

Next, it computes the minimum weight for each user to meet the above requirements, denoted by $\underline{w}_u$.  
If $l_b^{-v} + \underline{f}^v_b \le 1$,  the GREET resource allocation is given by 
\eqref{underload-allocation}, and slice $v$'s minimum local bid at base station $b$, $\underline{l}^v_b$, should satisfy 
$
\frac{\underline{l}^v_b}{\underline{l}^v_b  +{l}^{-v}_b } =  \underline{f}^v_b.
$
Hence, the minimum share for user $u$ at base station $b$ is given by  
$
\underline{w}_u = \frac{\underline{f}_u}{\underline{f}^v_b} \underline{l}^v_b 
= \frac{\underline{f}_u}{1- \underline{f}^v_b} {l}^{-v}_b
$
(Line 4).

If $l_b^{-v} + \underline{f}^v_b > 1$, the GREET resource allocation is given by \eqref{overload-allocation} and two cases
need to be considered. In first case, where the minimum resource allocation satisfies $\underline{f}^v_b  \leq s^v_b$, it suffices to
set  
$
\underline{l}^v_b = \underline{f}^v_b ~\mbox{and}~ \underline{w}_u = \underline{f}_u
$
and GREET resource allocation will make sure the requirement is met (Line 6).
In the second case, where $\underline{f}^v_b  >  s^v_b$, in order to meet
the minimal rate requirements under the GREET allocation given by \eqref{overload-allocation}, the minimum local bid
 allocation $\underline{l}^v_b$  must satisfy
$$
s^v_b + \frac{(\underline{l}^v_b - s^v_b)\left( 
1 - s^v_b - \sum\limits_{v'\in\mathcal{V}_b\setminus\{v\}} \min\left( s^{v'}_b, l^{v'}_b \right) \right)}
{ \underline{l}^v_b - s^v_b +\sum\limits_{v' \in \mathcal{V}_b\setminus\{v\}}\Delta^{v^\prime}_b} = \underline{f}^v_b.
$$
Solving the above for $\underline{l}^v_b$ and allocating user weights in proportion to $\underline{f}_u$ gives
the following minimum weights (Line~7): 
\begin{equation}
\underline{w}_u = \frac{\underline{f}_u}{\underline{f}^v_{b}} \bigg(s_b^v + 
\frac{(\underline{f}^v_b-s^v_b)\sum_{v' \in \mathcal{V}_b\setminus\{v\}}\Delta^{v^\prime}_b}{1 - \underline{f}^v_b - \sum_{v'\in\mathcal{V}_b\setminus\{v\}} \min( s^{v'}_b, l^{v'}_b)}\bigg).
\label{eq:weight-insufficient-guarantee}
\end{equation}						   		


Once we have computed the minimum weight requirement for all users, we proceed as follows. If the slice's overall share 
$s^v$ suffices to meet the requirements of all users, we divide the remaining share among the slice's users proportionally to their $\phi_u$ (Line 10). Otherwise, we assign weights such that we maximize the number of users that see their minimum rate requirement met, selecting users in order of increasing $\underline{w}_u$ and providing them with the minimum weight $\underline{w}_u$ (Lines 13-14). 


The lemma below lends support to the GREET share allocation algorithm. It shows that, 
under some relevant scenarios,
this algorithm captures the character of social optimal slice allocations.
Furthermore, 
in a network with many slices where the overall share of an individual slice is very small in relative terms,
GREET is a good approximation to a slice's best response, suggesting that a slice cannot gain (substantially) by deviating 
from GREET. This result thus confirms that, in addition to being simple, 
GREET provides close to optimal performance both at a global level 
(across the whole network) as well as locally (for each individual slice). 

\begin{lemma}\label{thm-approx}
The weight allocations provided by the GREET share allocation policy satisfy the following properties:
\begin{enumerate} 
\item Suppose that the users of all slices are elastic. Then, GREET provides all users with the same rate allocation as the social optimal weights, i.e., 
$
r_u(\mathbf{w}^{g}) = r_u(\mathbf{w}^{\textrm{so}}), \forall u,
$
where $\mathbf{w}^{\textrm{so}}$ is the (not necessarily unique) social optimal weight allocation and $\mathbf{w}^{g}$ is the weight allocation under GREET.
\item Suppose that all the users of a slice are either elastic or inelastic and Assumption 1 holds. Further, suppose that $s^v/l_b^{-v} < \epsilon \ \forall b$. Then, the following holds for all $u$:
\begin{equation*}
\frac{w_u^{br,v}(\mathbf{w}^{-v})}{1+\epsilon} < w_u^{g,v}(\mathbf{w}^{-v})
 < (1+\epsilon)w_u^{br,v}(\mathbf{w}^{-v}),
\end{equation*}
where $\mathbf{w}^{br,v}(\mathbf{w}^{-v})$ is the best response of slice $v$ to the other slices' weights $\mathbf{w}^{-v}$ and $\mathbf{w}^{g,v}(\mathbf{w}^{-v})$ is slice $v$'s response under GREET.

\end{enumerate}
\end{lemma}

One of the main goals of the GREET resource allocation model proposed in Section~\ref{sec-model}, in combination with the GREET share allocation policy proposed in this section, is to 
provide guarantees to different slices, so that they can in turn 
ensure that the minimum rate requirements of their users are met. 
The lemma below confirms that, as long as slices are well dimensioned, 
GREET will achieve this goal. 


\begin{lemma}\label{thm-min2}
When Assumption 1 holds, the resource allocation resulting from combining the GREET resource allocation model with the GREET share allocation policy
meets all users' minimum rate requirements.
\end{lemma}

\subsection{Convergence of the GREET algorithm}

A key desirable property for a slice-based share allocation policy is convergence 
to an equilibrium. Applying a similar argument to that of Theorem~\ref{thm-conv},
it can be shown that the GREET share allocation algorithm need not converge. 
However, below we will show sufficient conditions for convergence. 



We let  $\mathbf{w}(n)$ be the overall weight allocation for update round $n$. 
Our goal is to show that the weight sequence $\mathbf{w}(n)$ converges when $n \to \infty$. 
The following theorem provides a sufficient condition for geometric convergence
to a unique equilibrium. 
According to the theorem, convergence is guaranteed as long as 
($i$) slice shares are well dimensioned, 
and ($ii$) the guaranteed fraction of resources for a given slice at any base station is limited.
The second condition essentially says there should be quite a bit of flexibility
when managing guaranteed resources, leaving sufficient resources not committed to any slice. In practice, this may indeed
make sense in networks supporting slices with elastic traffic (which need non-committed resources), inelastic traffic (which may require some safety margins), or combinations thereof. 

\begin{theorem}\label{thm:sync-bid-convergence}
Suppose that Assumption 1 holds and the maximum aggregate resource requirement per slice, $f_{\max}$, satisfies 
\begin{equation}\label{eq:cond-convergence}
f_{\max} := \max_{v\in\mathcal{V}}\max_{b\in\mathcal{B}}\underline{f}_b^v
 < \frac{1} {2 |\mathcal{V}| -1}.
\end{equation}
Then, if slices perform GREET-based updates of their share allocations according to Algorithm 1,
either in Round Robin manner or simultaneously, 
the sequence of weight vectors $(\mathbf{w}(n):n\in\mathbb{N})$ converges to a unique 
fixed point, denoted by $\mathbf{w}^*$, irrespective of the initial share allocation $\mathbf{w}(0).$
Furthermore the convergence is geometric, i.e.,
\begin{eqnarray}\label{eq:norm-def}
\max_{v\in\mathcal{V}}\sum_{b\in\mathcal{B}}
|l^v_b(n) - l^{v,*}_b| \le
 \xi ^n \max_{v\in\mathcal{V}}\sum_{b\in\mathcal{B}}
|l^v_b(0) - l^{v,*}_b|
\end{eqnarray}
where $\xi := \frac{2 (|\mathcal{V}|-1)f_{\max}}{1 - f_{\max}}$ and $\mathbf{l}^{v,*}$ corresponds to slice v's per resource local bid at the fixed point  $\mathbf{w}^*$.
Note that, by \eqref{eq:cond-convergence}, we have $\xi < 1.$
\end{theorem}

This convergence result can be further generalized under the asynchronous update
model in continuous time~\cite{BeT89}. 
Specifically, without loss of generality, let $n$ index the sequence
of times $(t_n, n \in \mathbb{N})$ at which one or more slices update their share allocations
and let $\mathcal{N}^v$ denote the subset of those indices where slice $v$ performs an update. 
For $n \in \mathcal{N}^v$, slice $v$ updates its share allocations based on possibly
outdated weights for other slices, denoted by $(\mathbf{w}^{v'}(\tau_{v'}^v(n)) : v' \ne v)$, 
where $0 \le \tau_{v'}^v(n) \le n$ indexes the update associated with the
most recent slice $v'$ share weight updates available to slice $v$ prior to the $n^{th}$ update. 
As long as the updates are performed according to the assumption below, one can show that GREET converges under such asynchronous updates.

\begin{assumption}({\bf {\em Asynchronous updates}})
\label{asynchronous-assumption}
We assume that asynchronous updates are performed such that, for each slice $v \in {\cal V}$, the update sequence satisfies 
($i$) $|\mathcal{N}^v| = \infty$, and ($ii$) for any subsequence  $\{n_k\} \subset \mathcal{N}^v$ 
that tends to infinity, then $\lim_{k\rightarrow \infty} \tau^v_{v'}(n_k) = \infty, \ \forall v' \in \mathcal{V}$. 
\end{assumption}





\begin{theorem}\label{thm:async-convergence}
Under Assumption 1, if slices perform GREET-based updates of their share allocations 
asynchronously but satisfying Assumption \ref{asynchronous-assumption}, and 
if \eqref{eq:cond-convergence} holds, then the sequence of weight updates 
$(\mathbf{w}(n) : n\in\mathbb{N})$ converges to a unique fixed 
point irrespective of the initial condition.
\end{theorem}



While the above results provide some sufficient conditions for convergence, in the simulations performed in Section~\ref{sec-setup} we observed that, beyond these sufficient conditions, the algorithm always converges quite quickly under normal circumstances (within a few rounds). Based on this, we adopt an approach for the GREET share allocation algorithm where we let the weights to be updated by each slice for a number of rounds, and stop the algorithm if it has not converged upon reaching this number (which is set to 7 in our simulations).


%
%
%
%
%

\section{Performance Evaluation}\label{sec-setup}

In this section we present a detailed performance evaluation 
of GREET versus two representative slice-based resource allocation
approaches in the literature: one reservation- and the other share-based. 

\subsection{Mobile Network Simulation Setup}
\label{sec:simulation-real}

{\em Simulation model}: 
We simulate a dense `small cell' wireless deployment 
following the IMT-Advanced evaluation guidelines~\cite{IMT09}. The network 
consists of 19 base stations in a hexagonal cell layout with an inter-site distance of 20 meters and 3 sector antennas; thus, ${\cal B}$ corresponds to 57 sectors. Users associate to the sector offering the strongest SINR{\color{black},} where the downlink SINR between base station $b$ and user $u$ is modeled as in \cite{Ye13}:
$
\mbox{SINR}_{bu}=\frac{P_b G_{bu}}{\sum_{k\in \mathcal{B}\setminus\{b\}} P_k G_{ku} + \sigma^2},
$
where, following \cite{IMT09}, the noise $\sigma^2$ is set to $-104$dB, the transmit power $P_b$ is equal to $41$dB and the channel gain between BS sector $b$ and user $u$, denoted by $G_{bu}$, accounts for path loss, shadowing, fast fading and antenna gain. The path loss is defined as $36.7\log_{10}
(d_{bu})+22.7+26\log_{10}(f_c)$dB, where $d_{bu}$ denotes the current distance in meters from the user 
$u$ to sector $b$, and the carrier frequency $f_c$ is equal to $2.5$GHz. The antenna gain is set to 17 dBi, shadowing is updated every second and modeled by a log-normal distribution with standard deviation of 8dB~\cite{Ye13}; and fast fading follows a Rayleigh distribution depending on the mobile's speed and the angle of incidence. The achievable rate $c_{u}$ for user $u$ at a given point in time is based on a 
discrete set of modulation and coding schemes (MCS), with the associated SINR thresholds given in \cite{TS36213}. This MCS value is selected based on the average $\overline{\mbox{SINR}}_{bu}$, 
where channel fast fading is averaged over a second.
For user scheduling, we assume that resource blocks are assigned to users 
in a round-robin manner proportionally to the allocation determined by the resource allocation
policy under consideration.
For user mobility, we consider two different mobility patterns: Random Waypoint model (RWP)~\cite{HLV06}, yielding roughly uniform load distributions, and SLAW model \cite{slaw}, typically yielding clustered users and thus non-uniform load distributions.








\begin{figure*}[t]
\centering
  \begin{minipage}{.32\textwidth}
	\includegraphics[width = 1.05\textwidth]{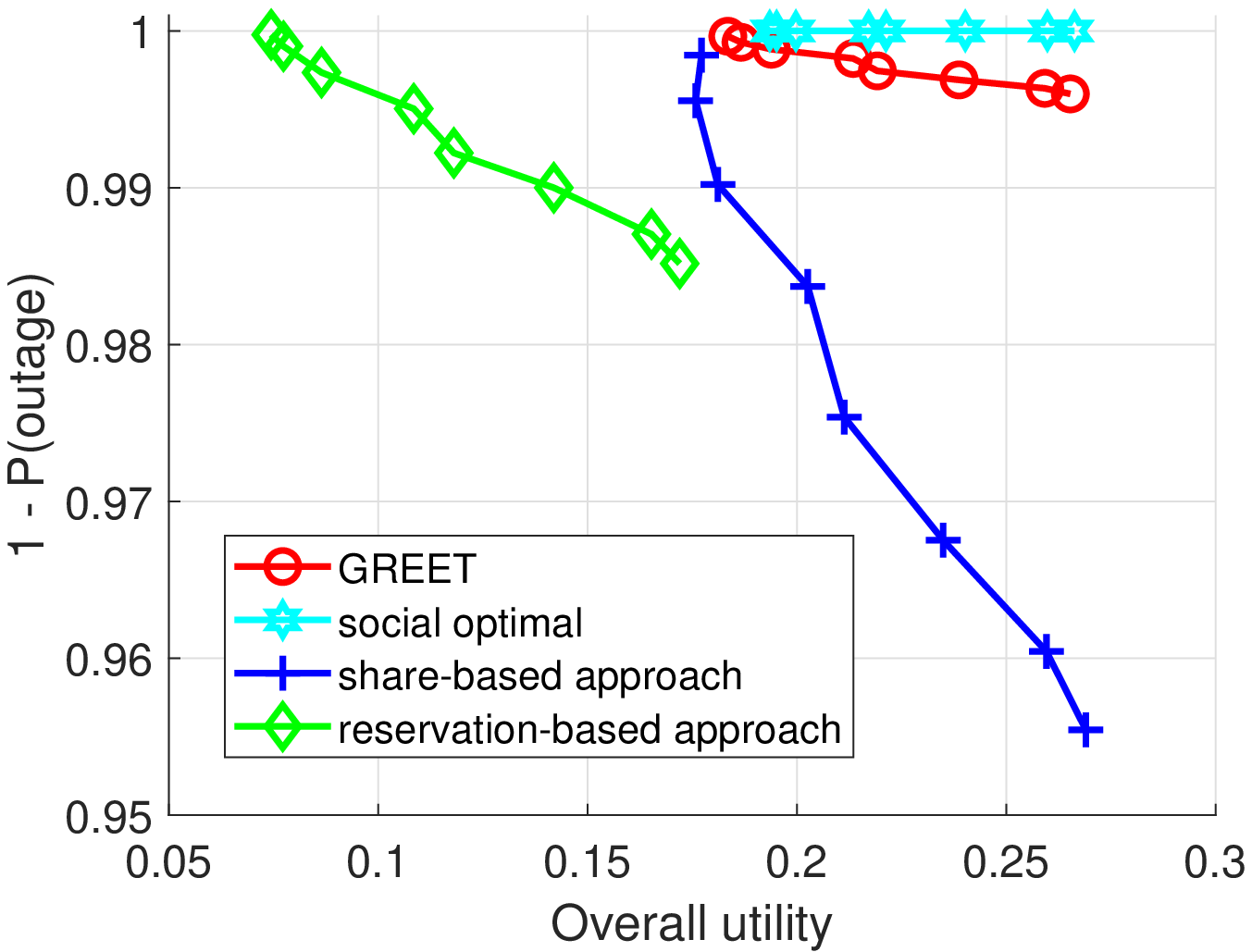}
	\caption{Comparison of GREET against the benchmark approaches in terms of the overall Utility $U$ and the outage probability $P(\mbox{outage})$.}
	\label{fig:util-outage-tradeoff-real}
	\end{minipage}\hfill
  \begin{minipage}{.32\textwidth}
	\includegraphics[width = 1.05\textwidth]{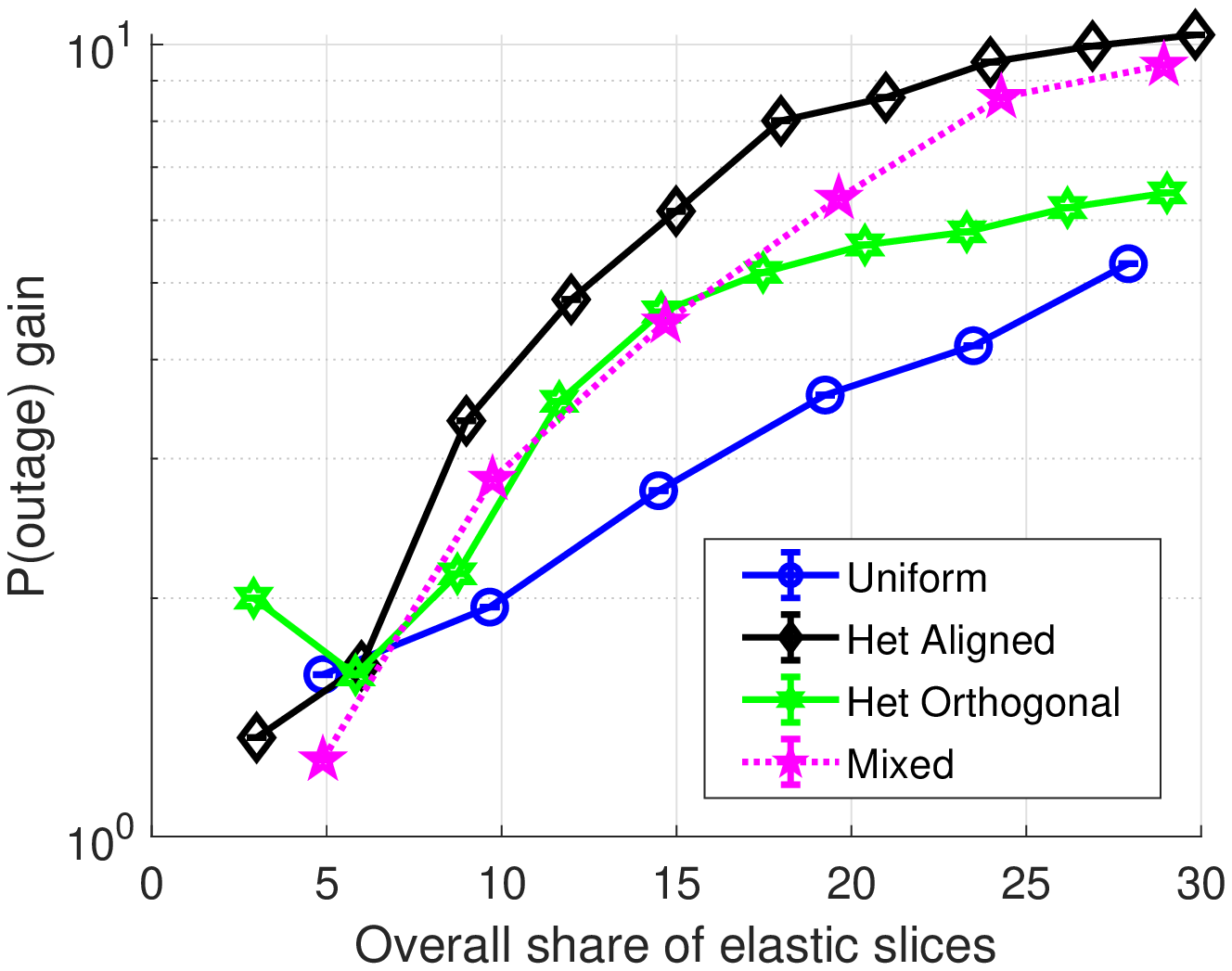}
	\caption{Gain in $P(\mbox{outage})$ over share-based approach, measured as the ratio of $P(\mbox{outage})$ under the share-based approach over that under GREET.}
	\label{fig:poutage-multicases}
	\end{minipage}\hfill
  \begin{minipage}{.32\textwidth}
	\includegraphics[width = 1.05\textwidth]{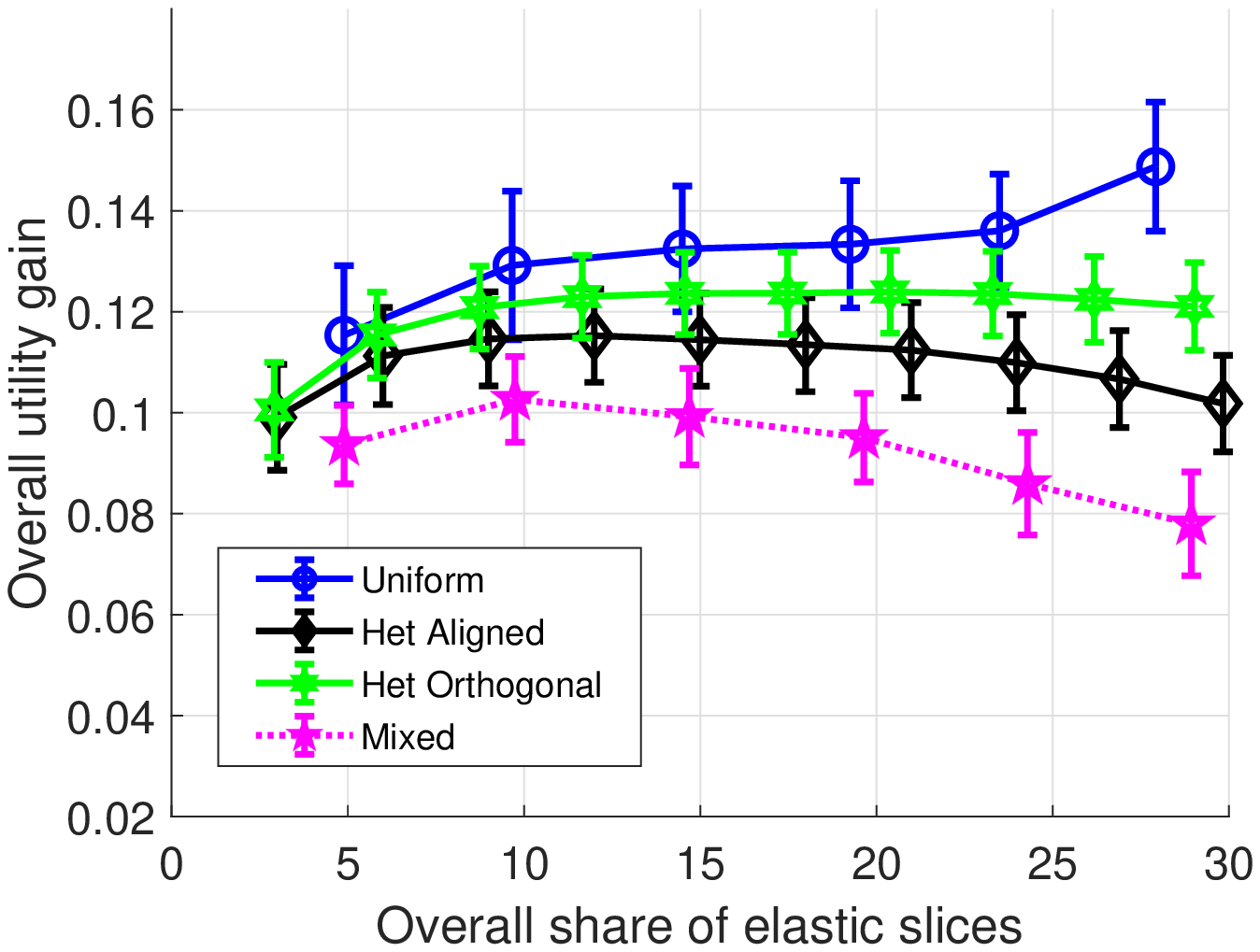}
	\caption{Gain in utility over reservation-based approach, measured as the utility under GREET minus that under the reservation-based scheme.}
	\label{fig:util-multicases}
	\end{minipage}%
\vspace{-0,25cm}
\end{figure*}

\vspace{0,125 cm}{\em Performance metrics}: 
Recall that our primary goal is 
to give slices flexibility in meeting their users' minimum rate requirements 
while optimizing the overall network efficiency. 
To assess the effectiveness of GREET in achieving this goal, we focus on the following two metrics:
\begin{itemize}[noitemsep,nolistsep]
\item Outage probability $P(\mbox{outage})$: this is the probability 
that a user does not meet its minimum rate requirement. 
In order for a slice to provide a reliable service, this probability should be kept below a certain threshold. 
\item Overall utility $U$: this is given by \eqref{eq:def-social-utility} and reflects the overall performance across all slices. 
\end{itemize}

\vspace{0,125 cm}{\em State-of-the-art approaches}: 
In order to show the advantages of GREET,
we will compare it to the following benchmarks:
\begin{itemize}[noitemsep,nolistsep]
\item Reservation-based approach: with this approach, each slice $v$ reserves a local share at each base station $b$, denoted
by $\hat{s}^{v}_b$. The resources at each base station are then shared among the {\em active} slices (having at least one user)
in proportion to the local shares $\hat{s}^{v}_b.$ This is akin to setting weights for a Generalized Processor Sharing in a resource~\cite{gps} and is in line with the spirit of reservation-based schemes in the literature~\cite{aanetworkslicing,globe17,Sci17,huawei,deepcog,Beg17,tmc}.
\item Share-based approach: with this approach, each slice gets a share $\tilde{s}^v$ of the 
overall resources, as in~\cite{ZCVB17,ZCD18,GPSB15,infocom17,ton,twcg}. Specifically, resources at each base station are shared according to SCPF as proposed in~\cite{ZCD18}, whereby each slice $v\in\mathcal{V}$ distributes its share $\tilde{s}^v$ equally amongst all its active users $u\in\mathcal{U}^v$, such that each user $u$ gets a weight $\tilde{w}_u = \tilde{s}^v/|\mathcal{U}^v|$, and then, at each base station $b\in\mathcal{B}$ the resources are allocated in proportion to users' weights.
\item Social optimal: this scheme corresponds to the social optimal
weight allocation $\mathbf{w}^{\textrm{so}}$ given by \eqref{eq:def-social-optimal}
under GREET resource allocation.
\end{itemize}





In order to meet the desired performance targets, the shares 
employed in the above approaches are dimensioned as follows. 
We consider two types of slices: ($i$) those which provide their users with
minimum rate requirements, which we refer to as \emph{guaranteed service slices}, 
and ($ii$) those which do not provide minimum rate requirements, 
which we refer to as \emph{elastic service slices}.
In GREET, for guaranteed service slices, we define a maximum acceptable 
outage probability $P_{max}$ and determine the necessary share at each base station, 
$s_b^v$, such that $P(\mbox{outage}) \le P_{max}$, assuming that the number of users 
follow a Poisson distribution whose mean is obtained from the simulated 
user traces; for these slices, we set $e^v = 0$. 
For elastic service slices, we set $s_b^v = 0 \ \forall b$ and $e^v$ to a value 
that determines the mean rate provided to elastic users. For the reservation-based approach, we set $\hat{s}_b^v = s_b^v$ for guaranteed service slices, to provide the same guarantees as GREET; for elastic service slices, we set $\hat{s}_b^v$ such that ($i$) their sum is equal to $e^v$, to provide the same total share as GREET, ($ii$) the sum of the $\hat{s}_b^v$'s at each base station does not exceed 1, to preserve the desired service guarantees, and ($iii$) they are as much balanced as possible across all base stations, within these two constraints. Finally, for the share-based approach we set $\tilde{s}^v = s^v$ for all slice types, i.e., the same shares as GREET.

\subsection{Comparison with state-of-the-art benchmarks}

Fig.~\ref{fig:util-outage-tradeoff-real} exhibits the performance of GREET versus the above 
benchmarks in terms of $P(\mbox{outage})$ and overall utility $U$ for the following scenario: 
($i$) we have two guaranteed service and two elastic service slices; 
($ii$) the share of elastic service slices is increased within the range $s^v \in [2,19]$;
($iii$) the minimum rate requirement for users on the guaranteed service slices is set
to $\gamma_u = 0.2 \, \mbox{Mbps} \ \forall u$; 
($iv$) the shares of guaranteed service slices
 are dimensioned to satisfy an outage probability threshold $P_{max}$ of $0.01$; 
($v$) for all slices, the priorities $\phi_u$ of all users are equal; 
and, ($vi$) the users of the elastic service slices follow 
the RWP model, leading to roughly uniform spatial loads, while the users of the guaranteed 
service slices have non-uniform loads as given by the SLAW model.
Since user utilities are not defined below the minimum rate requirements, 
the computation of the overall utility only takes into account the users whose 
minimum rate requirements are satisfied under all schemes. 

The results show that GREET outperforms both the 
share- and reservation-based approaches. 
While the share-based approach can flexibly shift resources across base stations, leading
to a good overall utility, it is not able to sufficiently isolate 
slices from one another, resulting in large outage probabilities, $P(\mbox{outage})$,
as the share of elastic service slices increase. 
By contrast, the reservation-based approach 
is effective in keeping $P(\mbox{outage})$ 
under control (albeit a bit above the threshold 
due to the approximation in the computation of $s_b^v$). 
However, since it relies on local decisions, it cannot globally optimize 
allocations and is penalized in terms of the overall utility. GREET achieves 
the best of both worlds: it meets the service requirements,
 keeping $P(\mbox{outage})$ well below the $P_{max}$ threshold, 
while achieving a utility that matches that of the share-based approach. 
Moreover, it performs very close to the social optimal, 
albeit with somewhat larger $P(\mbox{outage})$ due to the fact that the social optimal imposes the minimum rate requirements as constraints, forcing each slice to help the others meeting their minimum rate requirements, while in GREET each slice behaves `selfishly'.

\subsection{Outage probability gains over the share-based scheme}

One of the main observations of the experiment conducted above is 
that GREET provides substantial gains in terms of outage probability 
over the shared-based scheme. In order to obtain additional insights 
on these gains, we analyze them for a variety of scenarios 
comprising the following settings:
\begin{itemize}[noitemsep,nolistsep]
\item \emph{Uniform}: we have two guaranteed service slices 
and two elastic service slices; the users' mobility on all slices follow 
the RWP model and have the same priority $\phi_u$.
\item \emph{Heterogeneous Aligned}: 
the users of all slices are distributed non-uniformly according to SLAW but they all follow the same distribution 
(i.e., has same hotspots).
\item \emph{Heterogeneous Orthogonal}: 
all slices are distributed according to SLAW model but each slice follows a different distribution 
(i.e., has different hotspots). 
\item \emph{Mixed}: we have the same scenario as in Fig. \ref{fig:util-outage-tradeoff-real}, 
with the only difference that for one of the guaranteed service 
slices we have that all users are inelastic, i.e., the priority 
$\phi_u$ of all of them is set to 0.
\end{itemize}

For the above network configurations, we vary the share $s^v$ of elastic service slices while 
keeping the shares for the guaranteed service slices fixed. Fig.~\ref{fig:poutage-multicases} shows 
the ratio of the $P(\mbox{outage})$ of the share-based approach over 
that of GREET as a function of the overall share of elastic slices, i.e., $\sum_{v \in \mathcal{V}_e}{s^v}$, where $\mathcal{V}_e$ is the set of elastic service slices. Results are given with  95\% confidence intervals but they are so small that can barely be seen. We observe that GREET outperforms the 
share-based approach in all cases, providing $P(\mbox{outage})$ 
values up to one order of magnitude smaller. As expected, 
the gain in $P(\mbox{outage})$ grows as the the share of elastic service slices increases; indeed, 
as the share-based approach does not provide resource guarantees, it cannot control the outage probability of guaranteed service slices. 

\subsection{Utility gains over the reservation-based scheme}

In order to gain additional insight on the utility gains over 
the reservation-based scheme, in Fig.~\ref{fig:util-multicases} 
we analyze them for the scenarios introduced above. Results show that 
GREET consistently outperforms the reservation-based scheme across all approaches and share configurations, 
achieving similar gains in terms of overall utility in all cases.
This confirms that, by providing the ability to dynamically adjust 
the overall resource allocation to the current user distribution 
across base stations, GREET can achieve significant utility gains over the reservation-based approach. 


\section{Conclusions}\label{sec-conclusions}

GREET provides a flexible framework for managing heterogeneous performance requirements for network slices supporting dynamic user populations on a shared infrastructure. It is a practical approach that provides slices with sufficient resource guarantees to meet their requirements, and at the same time it allows them to unilaterally and dynamically customize their allocations to their current users' needs, thus achieving a good tradeoff between isolation and overall network efficiency. We view the GREET approach proposed here as a component of the overall solution to network slicing. Such a solution should include interfaces linking the resource allocation policies proposed here to lower level resource schedulers, which may possibly be opportunistic and delay-sensitive. Of particular interest will be the interfaces geared at supporting ultra-high reliability and with ultra-low latency services. 


\ifappendix
\section*{Appendix: Proofs of the Theorems} 

\ifextended
\subsection*{Proof of Lemma \ref{lm-convex}}
\else
\subsubsection*{Sketch of proof of Lemma \ref{lm-convex}}
\fi
\ifextended
We first show that there exists a weight setting that meets the minimum rate requirements of all users. As long as a fraction of base station $b$ equal to $s^v_b$ is sufficient to meet the user minimum rate requirements, by applying an aggregate weight equal to $s^v_b$ in the resource, the tenant is guaranteed to get this fraction of resources. As this can be applied to all resources, the minimum rate guarantees can be met for all the users of the tenant.

The optimization problem is given by the maximization of the sum of user utilities. This is a concave function on the weights as long as the individual user utilities are concave. As long as the minimum rate requirements are satisfied, individual user utilities are concave, as they are increasing concave function of a concave function (see~\cite{convexopt}). The set of feasible weights need to satisfy $\sum_{u \in \mathcal{U}^v}{w_u} \le s^v$ and $w_u \ge 0, \forall u\in\mathcal{U}$, and need to be such that the minimum rate requirements are satisfied. The latter imposes $w_u/\sum_{u \in \mathcal{U}_b}{w_u} \ge \gamma_u$ which yields $w_u - \gamma_u \sum_{u \in \mathcal{U}_b}{w_u} \ge 0, \ \forall u$. As a result, the set of flexible weights is convex.
\else
By allocating a local-bid equal to $s^v_b$ at base station $b$, slice $v$ 
is guaranteed a fraction $s^v_b$ of $b$'s resources. 
Under Assumption~\ref{assm:well-dimension}, this is sufficient to meet
the minimum rate guarantees. 
The convexity of the optimization follows from that of the utility and the feasible sets.
\fi
\ifextended
\subsection*{Proof of Theorem \ref{thm:ne-existence}}

We first prove the existence of a NE when $w_u \ge \delta$. Let $\mathcal{W}$ be the convex and compact set of feasible weights $\mathbf{w}$ satisfying ($i$) $w_u \ge \delta, \forall u$, and ($ii$) $\sum_{u \in \mathcal{U}^v}{w_u} = s^v, \forall v\in\mathcal{V}$ and let us consider the mapping $\mathbf{w} \to \tilde{\mathbf{w}} = \Gamma(\mathbf{w})$, where $\tilde{\mathbf{w}}^v$ is the best response of slice $v$ to $\mathbf{w}^{-v}$. We next show that this mapping satisfies the conditions of Kakutani's theorem: $i$) $\Gamma(\mathbf{w})$ is non-empty, $ii$) $\Gamma(\mathbf{w})$ is a convex-valued correspondence, and $iii$) $\Gamma(\mathbf{w})$ has a closed graph.

Conditions $i$) and $ii$) follow from Lemma \ref{lm-convex}.
According to this lemma, the best response of a slice to $\mathbf{w}^{-v}$ is the 
allocation $\tilde{\mathbf{w}}^v$ corresponding to the solution to a convex optimization problem. 
This implies that that $\tilde{\mathbf{w}}^v$ exists and is a convex set, 
according to the properties of convex optimization problems.
Hence, $\tilde{\mathbf{w}}$ exists and is a convex set as well.

Condition $iii$) is shown by proving that $\tilde{\mathbf{w}}^v$ is a continuous function of $\mathbf{w}^{-v}$ for all slices. Consider the set of base stations for which: ($i$) $l_b \le 1$, ($ii$) $l_b > 1$ and $f^v_b \leq s^v_b$, and ($iii$) $l_b > 1$ and $f^v_b > s^v_b$. As long as these sets do not change, $\tilde{\mathbf{w}}^v$ can be expressed as a continuously differentiable function of $\{\tilde{\mathbf{w}}^v,\mathbf{w}^{-v}\}$, and it follows from the implicit function theorem that $\tilde{\mathbf{w}}^v$ is a continuous function of $\mathbf{w}^{-v}$. When some base station moves from one set to the other, such base station satisfies the equations corresponding to both cases, providing continuity over the transitions.

Since all the conditions of Kakutani's theorem are satisfied, we have that the mapping $\Gamma$ has at least one fixed point, which implies that at least one NE exists.

We now prove that $w_u \ge \delta$ does not hold, we may not have a NE. Consider a scenario with two slices, 1 and 2, and two base stations, a and b. Each slice has a user in each base station such that $\gamma_{1a} = \gamma_{2a} = 1/4$ and $\gamma_{1b} = \gamma_{2b} = 0$. Furthermore, we have $\phi_{1a} = \phi_{2a} = 0$, $\phi_{1b} = \phi_{2b} = 1$, $s^1_a = s^2_a = 1$, $s^1_b = s^2_b = 0$ and $e^1 = e^2 = 0$. In the best response, it holds $w_{1a} = w_{1b}/3$ and $w_{1b} = w_{1a}/3$, which implies that there exists no NE.
\else
\vspace{0,2cm}
\subsubsection*{Sketch of the proof of Theorem \ref{thm:ne-existence}}

We first prove the existence of a NE when $w_u \ge \delta$. Let $\mathcal{W}$ be the convex and compact set of feasible weights $\mathbf{w}$ satisfying ($i$) $w_u \ge \delta, \forall u$, and ($ii$) $\sum_{u \in \mathcal{U}^v}{w_u} = s^v, \forall v\in\mathcal{V}$ and let us consider the mapping $\mathbf{w} \to \tilde{\mathbf{w}} = \Gamma(\mathbf{w})$, where $\tilde{\mathbf{w}}^v$ is the best response of slice $v$ to $\mathbf{w}^{-v}$. We show that this mapping satisfies the conditions of Kakutani's theorem: $i$) $\Gamma(\mathbf{w})$ is non-empty, $ii$) $\Gamma(\mathbf{w})$ is a convex-valued correspondence, and $iii$) $\Gamma(\mathbf{w})$ has a closed graph. Conditions $i$) and $ii$) follow from Lemma \ref{lm-convex} and the properties of convex optimization problems, which ensure that $\tilde{\mathbf{w}}^v$ exists and is a convex set. Condition $iii$) is shown by proving that $\tilde{\mathbf{w}}^v$ is a continuous function of $\mathbf{w}^{-v}$ for all slices. Since all the conditions of Kakutani's theorem are satisfied, we have that the mapping $\Gamma$ has at least one fixed point, which implies that at least one NE exists. To prove that if $w_u \ge \delta$ does not hold, we may not have a NE, we consider a scenario with two slices, 1 and 2, and two base stations, $a$ and $b$. Each slice has a user in each base station such that $\gamma_{1a} = \gamma_{2a} = 1/4$ and $\gamma_{1b} = \gamma_{2b} = 0$. Furthermore, we have $\phi_{1a} = \phi_{2a} = 0$, $\phi_{1b} = \phi_{2b} = 1$, $s^1_a = s^2_a = 1$, $s^1_b = s^2_b = 0$ and $e^1 = e^2 = 0$. In the best response, it needs to hold $w_{1a} = w_{1b}/3$ and $w_{1b} = w_{1a}/3$. As these cannot be satisfied simultaneously, there exists no NE.
\fi

\ifextended
\subsection*{Proof of Theorem \ref{thm-conv}}

Let us consider a scenario with three slices, denoted by Slices 1, 2 and 3, and three base stations, denoted by Base Station (BS) a, b, and c, respectively.
Slice 1 has two users, one at BS a, another at BS b, denoted by 
1a and 1b, respectively. Slice 2 has two users, one at BS b, another at BS c, denoted by 2b and 2c. Also, Slice 3 has two users at BS a and c, respectively, denoted by 3a and 3c.
The share allocation is $s^1 = s^2 = s^3 = 3/4 + \epsilon$ for some $\delta < \epsilon < 1/4$, $s^1_a = s^2_b  = s^3_c = 3/4$, $\gamma_{1a} = \gamma_{2b} = \gamma_{3c} = 3/4$, $\gamma_{1b} = \gamma_{2c} = \gamma_{3a} = 0$, $\phi_{1a} = \phi_{2b} = \phi_{3c} = 0$ and $\phi_{1b} = \phi_{2c} = \phi_{3a} = 1$. 

It can be seen that a NE in the above scenario is given by $w_{1a} = w_{2b} = w_{3c} = 9/16 + 3\epsilon/4$ and $w_{1b} = w_{2c} = w_{3a} = 3/16+\epsilon/4$.

Let us start with $w_{3a} > 1/4$ and and apply the best response starting with slice 1 followed by 2 and 3. Slice 1 takes $w_{1a} = 3/4$ and $w_{1b} = \epsilon$. In turn, slice 2 selects $w_{2b} = 3\epsilon$ and $w_{2c} = 3/4 - 2\epsilon > 1/4$. This yields $w_{3c} = 3/4$ and $w_{3a} = \epsilon$. We thus enter an endless cycle where $w_{1a}$, $w_{2b}$ and $w_{3c}$ alternate the values of $3/4$ with $3\epsilon$.
\else
\vspace{0,2cm}
\subsubsection*{Sketch of the proof of Theorem \ref{thm-conv}}

Let us consider a scenario with three slices, 1, 2 and 3, and three base stations (BS's), $a$, $b$, and $c$. Slice 1 has two users at BS's $a$ and $b$: users 1$a$ and 1$b$. Slice 2 has two users at BS's $b$ and $c$, 2$b$ and 2$c$, and slice 3 has two users at BS's $a$ and $c$, 3$a$ and 3$c$. The share allocation is $s^1 = s^2 = s^3 = 3/4 + \epsilon$ for some $\delta < \epsilon < 1/4$, $s^1_a = s^2_b  = s^3_c = 3/4$, $\gamma_{1a} = \gamma_{2b} = \gamma_{3c} = 3/4$, $\gamma_{1b} = \gamma_{2c} = \gamma_{3a} = 0$, $\phi_{1a} = \phi_{2b} = \phi_{3c} = 0$ and $\phi_{1b} = \phi_{2c} = \phi_{3a} = 1$. It can be seen that a NE in the above scenario is given by $w_{1a} = w_{2b} = w_{3c} = 9/16 + 3\epsilon/4$ and $w_{1b} = w_{2c} = w_{3a} = 3/16+\epsilon/4$. However, if we start with $w_{3a} > 1/4$ and apply the best response starting with slice 1 followed by 2 and 3, we enter an endless cycle where $w_{1a}$, $w_{2b}$ and $w_{3c}$ alternate the values of $3/4$ with $3\epsilon$.
\fi

\ifextended
\subsection*{Proof of Theorem \ref{thm-approx}}
By employing the KKT method with the constraint that all users should have a rate larger than or equal to the minimum rate  requirement, it can be seen that the optimal allocation is precisely the one that allocates the weight to all users according to the above. The analysis is similar to that in \cite{GPSB15} adding the minimum rate constraints. This allocation coincides with the one provided by the practical approach.

According to \cite{infocom17}, the best response is the weight allocation that assigns each user a weight proportional to $\frac{a_b^v}{a_b^v+l_b^v}$,
where $v$ denotes the slice that user belongs to and $b$ the base station it is associated with. 
When considering the minimum rate requirements, this imposes that for some users we may need a larger weight. The asymptotic case corresponds to the case where the number of tenants grows to infinite and thus the share of each individual tenant is negligible with respect to the sum of shares, i.e., $\frac{s_v}{\sum_{v^\prime}{s_{b}^{v^\prime}}}\to 0$. This leads to $\frac{a_b^o}{a_b^o+l_b^o} \to 1$. Consequently, this means that we assign each user the same weight (except for those that require a higher weight to meet their minimum rate requirement). This is precisely the allocation provided by the practical algorithm.
\else
\vspace{0,2cm}
\subsubsection*{Sketch of proof of Lemma \ref{thm-approx}}
The first statement is shown by solving the social optimal allocation following a similar argument to~\cite{GPSB15}. The second statement is shown by computing the best response similarly to~\cite{infocom17} and noting that $l_b^{br,v}(\mathbf{w}^{-v})/l_b^{-v} \le s^v/l_b^{-v} < \epsilon$.
\fi

\ifextended
\subsection*{Proof of Theorem \ref{thm:sync-bid-convergence}}

We show convergence of the slice shares by showing the Algorithm 1
is a contraction mapping. 
Specifically, consider two sequences slice-based share allocations
denoted $(\mathbf{l}(n): n\in\mathbb{N})$ and $(\tilde{\mathbf{l}}(n): n\in\mathbb{N})$, where
$\mathbf{l}(n) := (l^v_b(n) : v\in\mathcal{V}, b\in\mathcal{B})$
and $\tilde{\mathbf{l}}(n) := (\tilde{l}^v_b(n) : v\in\mathcal{V}, b\in\mathcal{B})$,
corresponding to two initial share allocations denoted 
denoted $\mathbf{l}(0),\tilde{\mathbf{l}}(0)$ where at each step each slice performs
its GREET share allocation in response to that of the others in the previous step.
We will establish that regardless the initial conditions 
$$
\max_{v\in\mathcal{V}}\sum_{b\in\mathcal{B}} |l^v_b(n) - \tilde{l}^v_b(n)| ~\le  
~\xi \max_{v\in\mathcal{V}}\sum_{b\in\mathcal{B}}
|l^v_b(n - 1) - \tilde{l}^v_b(n - 1)|
$$
which if $\xi < 1$ suffices to establish convergence.

We let $\underline{\mathbf{l}}(n) : = (\underline{l}^v_b(n) : 
v\in\mathcal{V}, b\in\mathcal{B})$ denote the minimal slice 
share allocations required by slice $v$ at base station $b$ based
on the share allocations in the previous round, i.e.,  $\mathbf{l}(n-1)$. 
Under Assumption 1, only Lines 4 and 5 in Algorithm 1 will
be in effect, so  
\begin{eqnarray}
\underline{l}_b^v(n) = \begin{cases}
\frac{\underline{f}^v_b}{1 - \underline{f}^v_b} l_b^{-v}(n - 1), & l^{-v}_b(n - 1) + \underline{f}^v_b
\le 1,   \\
\underline{f}^v_b, & l^{-v}_b(n - 1) + \underline{f}^v_b > 1.
\end{cases}
\label{eq:minimal-lob}
\end{eqnarray}
Again under Assumption 1 the share allocations for each 
slice and base station in response to the others 
${\mathbf{l}}(n)$ is given by Line 21 in Algorithm 1, i.e.,
$
l^v_b(n) =   
\underline{l}^v_b (n) + 
\phi^v_b \left(s^v - \sum_{b^\prime \in \mathcal{B}} \underline{l}^v_{b^\prime} (n)\right)
$
where $\phi^v_b = \sum_{u \in {\cal U}^v_b} \phi_u.$ 
Note if a slice $v$ has solely inelastic users $\phi^v_b= 0$ so 
$l^v_b(n) = \underline{l}^v_b (n).$ If a slice has solely elastic users then  
$\underline{l}^v_{b'} (n) = 0$ for all $b' \in {\cal B}$ and
$l^v_b(n) = \phi^v_b s^v$, while the general case is for customers with 
minimal rate constraints which are also rate adaptive. 
We define $\underline{\tilde{\mathbf{l}}}(n)$ in the same way based on  ${\tilde{\mathbf{l}}}(n).$ 

Next consider the difference between the two share allocation sequences which using
the Triangle inequality gives
\begin{equation*}
|l^v_b(n) - \tilde{l}^v_b(n)| \le 
|\underline{l}^v_b(n) - \underline{\tilde{l}}^v_b(n)|
+ \phi^b_v
\sum_{b^\prime \in \mathcal{B}} |
\underline{l}^v_{b^\prime}(n) - \underline{\tilde{l}}^v_{b^\prime}(n)|.
\end{equation*}

Noting that Eq.(\ref{eq:minimal-lob}) is a concave function 
with slope no greater than $\frac{\underline{f}^v_b}{1 - \underline{f}^v_b}$ and
again using the Triangle inequality we have that
\begin{eqnarray*}
|\underline{l}^v_b(n) - \underline{\tilde{l}}^v_b(n)| & \le & 
 \frac{\underline{f}^v_b}{1 - \underline{f}^v_b}|l^{-v}_b(n - 1) - \tilde{l}^{-v}_b(n - 1)| \\
&  \le & \frac{\underline{f}^v_b}{1 - \underline{f}^v_b}\sum_{v^\prime \ne v}
|l^{v^\prime}_b(n - 1) - \tilde{l}^{v^\prime}_b(n - 1)|.
\end{eqnarray*}

Thus, after one round of share updates, we have the following bound
\begin{eqnarray}
\lefteqn{
|l^v_b(n) - \tilde{l}^v_b(n)| \le \frac{\underline{f}^v_b}{1 - \underline{f}^v_b} \sum_{v^\prime \ne v}
\left| l^{v^\prime}_b(n - 1) - \tilde{l}^{v^\prime}_b(n - 1) \right|}\nonumber\\
& & + {\phi^v_b} \sum_{b^\prime \in \mathcal{B}}
\frac{\underline{f}^v_{b^\prime}}{1 - \underline{f}^v_{b^\prime}}
\sum_{v^\prime \ne v}\left| l^{v^\prime}_{b^\prime}(n - 1) - \tilde{l}^{v^\prime}_{b^\prime}(n - 1) \right|.
\label{eq:bound-single-diff}
\end{eqnarray}
This in turn leads to the following bound on $\mathbf{l}(n) - \mathbf{\tilde{l}}(n)$:
\begin{eqnarray*}
\lefteqn{\max_{v\in\mathcal{V}}\sum_{b\in\mathcal{B}}
|l^v_b(n) - \tilde{l}^v_b(n)|
} \\
& \le & \max_{v\in\mathcal{V}} \sum_{b\in\mathcal{B}}
\left\{ \frac{\underline{f}^v_b}{1 - \underline{f}^v_b} \sum_{v^\prime \ne v}
|l^{v^\prime}_b(n - 1) - \tilde{l}^{v^\prime}_b(n - 1)| \right. \\
& & + \left. {\phi^v_b} \sum_{b^\prime \in \mathcal{B}}
\frac{\underline{f}^v_{b^\prime}}{1 - \underline{f}^v_{b^\prime}} \sum_{v^\prime \ne v}
|l^{v^\prime}_{b^\prime}(n - 1) - \tilde{l}^{v^\prime}_{b^\prime}(n - 1)| \right\} \\
& \le & \frac{2 (|\mathcal{V}|-1){f}_{\max}}{1 - {f}_{\max}} 
\max_{v\in\mathcal{V}} \sum_{b\in\mathcal{B}}
|l^v_b(n - 1) - \tilde{l}^v_b(n - 1)|,
\end{eqnarray*} 
where we have used the bound $f_{\max}$ and that
$\sum_{b \in {\cal B}} \phi^v_b =1$ unless slice $v$ is inelastic in which case it equals 0.
If Eq. (\ref{eq:cond-convergence}) holds we have that
the share allocation updates get closer. 
It follows by Proposition 1.1 in Chapter 3 of \cite{BeT89} that under
simultaneous updates one has  geometric convergence to
fixed point. Similarly under round-robin updates, 
geometric convergence follows as a result 
of Proposition 1.4 in Chapter 3 of \cite{BeT89}.
\else
\vspace{0,2cm}
\subsubsection*{Proof of Theorem \ref{thm:sync-bid-convergence}}

We show convergence of the slice share allocation by showing that Algorithm 1
is a contraction mapping. 
Specifically, consider two share allocation sequences
denoted $(\mathbf{l}(n): n\in\mathbb{N})$ and $(\tilde{\mathbf{l}}(n): n\in\mathbb{N})$, where
$\mathbf{l}(n) := (l^v_b(n) : v\in\mathcal{V}, b\in\mathcal{B})$
and $\tilde{\mathbf{l}}(n) := (\tilde{l}^v_b(n) : v\in\mathcal{V}, b\in\mathcal{B})$,
corresponding to two initial share allocations $\mathbf{l}(0),\tilde{\mathbf{l}}(0)$, where at each step each slice computes
its GREET share allocation in response to the other slices' share allocations in the previous step.
%
We let $\underline{\mathbf{l}}(n) : = (\underline{l}^v_b(n) : 
v\in\mathcal{V}, b\in\mathcal{B})$ denote the minimal slice 
share allocation required by slice $v$ at base station $b$ based
on the share allocations in the previous round, i.e.,  $\mathbf{l}(n-1)$. 
Under Assumption 1, we have
\begin{eqnarray}
\underline{l}_b^v(n) = \begin{cases}
\frac{\underline{f}^v_b}{1 - \underline{f}^v_b} l_b^{-v}(n - 1), & l^{-v}_b(n - 1) + \underline{f}^v_b
\le 1,   \\
\underline{f}^v_b, & l^{-v}_b(n - 1) + \underline{f}^v_b > 1.
\end{cases}
\label{eq:minimal-lob}
\end{eqnarray}

The share allocations for each slice in response to the others is given by 
$
l^v_b(n) =   
\underline{l}^v_b (n) + 
\phi^v_b \left(s^v - \sum_{b^\prime \in \mathcal{B}} \underline{l}^v_{b^\prime} (n)\right),
$
where $\phi^v_b = \sum_{u \in {\cal U}^v_b} \phi_u.$ 
We define $\underline{\tilde{\mathbf{l}}}(n)$ in the same way based on  ${\tilde{\mathbf{l}}}(n).$ 
If we consider the difference between the two share allocation sequences and use
the Triangle inequality, this gives
\begin{equation*}
|l^v_b(n) - \tilde{l}^v_b(n)| \le 
|\underline{l}^v_b(n) - \underline{\tilde{l}}^v_b(n)|
+ \phi^b_v
\sum_{b^\prime \in \mathcal{B}} |
\underline{l}^v_{b^\prime}(n) - \underline{\tilde{l}}^v_{b^\prime}(n)|.
\end{equation*}

Noting that \eqref{eq:minimal-lob} is a concave function 
with slope no greater than $\frac{\underline{f}^v_b}{1 - \underline{f}^v_b}$ and using again the Triangle inequality, we have
\begin{eqnarray*}
|\underline{l}^v_b(n) - \underline{\tilde{l}}^v_b(n)| & \le & 
 \frac{\underline{f}^v_b}{1 - \underline{f}^v_b}|l^{-v}_b(n - 1) - \tilde{l}^{-v}_b(n - 1)| \\
&  \le & \frac{\underline{f}^v_b}{1 - \underline{f}^v_b}\sum_{v^\prime \ne v}
|l^{v^\prime}_b(n - 1) - \tilde{l}^{v^\prime}_b(n - 1)|.
\end{eqnarray*}

Thus, after one round of share updates, we have 
\begin{eqnarray*}
\lefteqn{
|l^v_b(n) - \tilde{l}^v_b(n)| \le \frac{\underline{f}^v_b}{1 - \underline{f}^v_b} \sum_{v^\prime \ne v}
\left| l^{v^\prime}_b(n - 1) - \tilde{l}^{v^\prime}_b(n - 1) \right|}\nonumber\\
& & + {\phi^v_b} \sum_{b^\prime \in \mathcal{B}}
\frac{\underline{f}^v_{b^\prime}}{1 - \underline{f}^v_{b^\prime}}
\sum_{v^\prime \ne v}\left| l^{v^\prime}_{b^\prime}(n - 1) - \tilde{l}^{v^\prime}_{b^\prime}(n - 1) \right|.
\label{eq:bound-single-diff}
\end{eqnarray*}
This in turn leads to the following bound on $\mathbf{l}(n) - \mathbf{\tilde{l}}(n)$:
\begin{eqnarray*}
\lefteqn{\max_{v\in\mathcal{V}}\sum_{b\in\mathcal{B}}
|l^v_b(n) - \tilde{l}^v_b(n)|
} \\
& \le & \max_{v\in\mathcal{V}} \sum_{b\in\mathcal{B}}
\left\{ \frac{\underline{f}^v_b}{1 - \underline{f}^v_b} \sum_{v^\prime \ne v}
|l^{v^\prime}_b(n - 1) - \tilde{l}^{v^\prime}_b(n - 1)| \right. \\
& & + \left. {\phi^v_b} \sum_{b^\prime \in \mathcal{B}}
\frac{\underline{f}^v_{b^\prime}}{1 - \underline{f}^v_{b^\prime}} \sum_{v^\prime \ne v}
|l^{v^\prime}_{b^\prime}(n - 1) - \tilde{l}^{v^\prime}_{b^\prime}(n - 1)| \right\} \\
& \le & \frac{2 (|\mathcal{V}|-1){f}_{\max}}{1 - {f}_{\max}} 
\max_{v\in\mathcal{V}} \sum_{b\in\mathcal{B}}
|l^v_b(n - 1) - \tilde{l}^v_b(n - 1)|,
\end{eqnarray*} 
where we have used the bound $f_{\max}$ and that
$\sum_{b \in {\cal B}} \phi^v_b =1$.
If \eqref{eq:cond-convergence} holds, we have that
the share allocation updates get closer. 
It follows by Proposition 1.1 in \cite[Chapter 3]{BeT89} that under
simultaneous updates one has  geometric convergence to
fixed point. Similarly, under round-robin updates, 
geometric convergence follows from Proposition 1.4 in \cite[Chapter 3]{BeT89}.
\fi

\ifextended
\subsection*{Proof of Theorem \ref{thm:async-convergence}}

This follows directly from the proof of Theorem~\ref{thm:sync-bid-convergence} and Proposition~2.1 in Chapter 6 of \cite{BeT89}.
\else
\vspace{0,2cm}
\subsubsection*{Sketch of the proof of Theorem \ref{thm:async-convergence}}

This follows from the proof of Theorem~\ref{thm:sync-bid-convergence} and Proposition~2.1 in~\cite[Chapter 6]{BeT89}.
\fi

\fi

\bibliographystyle{IEEEtran}
\bibliography{bibliography,slicing}

\begin{thebibliography}{10}
\providecommand{\url}[1]{#1}
\csname url@samestyle\endcsname
\providecommand{\newblock}{\relax}
\providecommand{\bibinfo}[2]{#2}
\providecommand{\BIBentrySTDinterwordspacing}{\spaceskip=0pt\relax}
\providecommand{\BIBentryALTinterwordstretchfactor}{4}
\providecommand{\BIBentryALTinterwordspacing}{\spaceskip=\fontdimen2\font plus
\BIBentryALTinterwordstretchfactor\fontdimen3\font minus
  \fontdimen4\font\relax}
\providecommand{\BIBforeignlanguage}[2]{{%
\expandafter\ifx\csname l@#1\endcsname\relax
\typeout{** WARNING: IEEEtran.bst: No hyphenation pattern has been}%
\typeout{** loaded for the language `#1'. Using the pattern for}%
\typeout{** the default language instead.}%
\else
\language=\csname l@#1\endcsname
\fi
#2}}
\providecommand{\BIBdecl}{\relax}
\BIBdecl

\bibitem{aanetworkslicing}
S.~Vassilaras and et. al., ``{The Algorithmic Aspects of Network Slicing},''
  \emph{IEEE Comm. Mag.}, vol.~55, no.~8, pp. 112--119, Aug. 2017.

\bibitem{globe17}
G.~W. et~al., ``{Resource Allocation for Network Slices in 5G with Network
  Resource Pricing},'' in \emph{Proc. IEEE GLOBECOM}, Dec. 2017.

\bibitem{Sci17}
V.~Sciancalepore and et. al., ``{Mobile Traffic Forecasting for Maximizing 5G
  Network resource Utilization},'' in \emph{Proc. IEEE INFOCOM}, May 2017.

\bibitem{huawei}
M.~Leconte and et. al., ``A resource allocation framework for network
  slicing,'' in \emph{Proc. IEEE INFOCOM}, Apr. 2018.

\bibitem{deepcog}
D.~Bega and et. al., ``Deepcog: Cognitive network management in sliced 5g
  networks with deep learning,'' in \emph{Proc. IEEE INFOCOM}, Apr. 2019.

\bibitem{Beg17}
------, ``{Optimising 5G Infrastructure Markets: The Business of Network
  Slicing},'' in \emph{Proc. IEEE INFOCOM}, May 2017.

\bibitem{tmc}
------, ``{Mobile Traffic Forecasting for Maximizing 5G Network resource
  Utilization},'' in \emph{IEEE Tran. Mobile Comp.}, to appear.

\bibitem{orion}
X.~Foukas and et. al., ``{Orion: RAN slicing for a flexible and cost-effective
  multi-service mobile network architecture},'' in \emph{Proc. ACM MOBICOM)},
  Oct. 2017.

\bibitem{ZCVB17}
J.~Zheng and et. al., ``Statistical multiplexing and traffic shaping games for
  network slicing,'' in \emph{Proc. WiOPT}, Paris, France, May 2017.

\bibitem{ZCD18}
------, ``Statistical multiplexing and traffic shaping games for network
  slicing,'' \emph{IEEE/ACM Trans. Networking}, vol.~26, no.~6, pp. 2528--2541,
  Dec 2018.

\bibitem{GPSB15}
P.~Caballero and et. al., ``{Multi-Tenant Radio Access Network Slicing:
  Statistical Multiplexing of Spatial Loads},'' \emph{IEEE/ACM Trans.
  Networking}, vol.~25, no.~5, pp. 3044--3058, Oct 2017.

\bibitem{infocom17}
------, ``{Network Slicing Games: Enabling Customization in Multi-Tenant
  Networks},'' in \emph{Proc. IEEE INFOCOM}, May 2017.

\bibitem{ton}
------, ``{Network Slicing Games: Enabling Customization in Multi-Tenant Mobile
  Networks},'' \emph{IEEE/ACM Trans. Networking}, vol.~27, no.~2, pp. 662--675,
  Apr. 2017.

\bibitem{twcg}
------, ``{Network Slicing for Guaranteed Rate Services: Admission Control and
  Resource Allocation Games},'' \emph{IEEE Tran. Wireless Comm.}, vol.~17,
  no.~10, pp. 6419--6432, Oct. 2018.

\bibitem{oroton}
S.~D'Oro, F.~Restuccia, T.~Melodia, and S.~Palazzo, ``{Low-complexity
  distributed radio access network slicing: Algorithms and experimental
  results},'' \emph{IEEE/ACM Transactions on Networking}, vol.~26, no.~6, pp.
  2815--2828, Dec. 2018.

\bibitem{son2}
{3GPP}, ``{Self-Organizing Networks (SON) Policy Network Resource Model (NRM)
  Integration Reference Point (IRP); Information Service (IS)},'' TS 28.628,
  Jun. 2013.

\bibitem{Mah13}
R.~Mahindra, M.~A. Khojastepour, H.~Zhang, and S.~Rangarajan, ``{Radio Access
  Network sharing in cellular networks},'' in \emph{Proc. of IEEE ICNP}, Oct.
  2013.

\bibitem{oroinfocom}
S.~D’Oro and et. al., ``{The Slice Is Served: Enforcing Radio Access Network
  Slicing in Virtualized 5G Systems},'' in \emph{Proc. IEEE INFOCOM}, Apr.
  2019.

\bibitem{mandelli}
S.~Mandelli and et. al., ``{Satisfying Network Slicing Constraints via 5G MAC
  Scheduling},'' in \emph{Proc. IEEE INFOCOM)}, Apr. 2019.

\bibitem{ksentini}
A.~Ksentini and N.~Nikaein, ``{Toward enforcing network slicing on RAN:
  Flexibility and resources abstraction},'' \emph{IEEE Comm. Mag.}, vol.~55,
  no.~6, pp. 102--108, Jun. 2017.

\bibitem{Ric16}
M.~Richart and et. al., ``Resource slicing in virtual wireless networks: A
  survey,'' \emph{IEEE Tran. Network Service Management}, vol.~13, no.~3, pp.
  462--476, Sept 2016.

\bibitem{AsM13}
A.~Asadi and V.~Mancuso, ``A survey on opportunistic scheduling in wireless
  communications,'' \emph{IEEE Comm. Surveys and Tutorials}, vol.~15, no.~4,
  pp. 1671--88, 2013.

\bibitem{borst2009mobility}
S.~Borst, N.~Hegde, and A.~Proutiere, ``{Mobility-Driven Scheduling in Wireless
  Networks},'' in \emph{Proc. IEEE INFOCOM}, Apr. 2009.

\bibitem{kushner2004conv}
H.~J. Kushner and P.~A. Whiting, ``Convergence of {P}roportional-{F}air
  {S}haring {A}lgorithms {U}nder {G}eneral {C}onditions,'' \emph{IEEE Trans.
  Wireless Comm.}, vol.~3, no.~4, pp. 1250--1259, Jul. 2004.

\bibitem{kalil2015qos}
M.~Kalil, A.~Shami, and A.~Al-Dweik, ``{QoS-Aware} {P}ower-{E}fficient
  {S}cheduler for {LTE} {U}plink,'' \emph{IEEE Trans. Mobile Comp.}, vol.~14,
  no.~8, pp. 1672--1685, Aug. 2015.

\bibitem{Mattess2010}
M.~Mattess, C.~Vecchiola, and R.~Buyya, ``{Mobility-Driven Scheduling in
  Wireless Networks},'' in \emph{Proc. IEEE HPCC}, Sep. 2010.

\bibitem{amazon}
Amazon, ``{EC2 Spot Instances},'' \url{https://aws.amazon.com/ec2/spot/}.

\bibitem{google}
{Google Cloud}, ``{Preemptible Virtual Machines},''
  \url{https://cloud.google.com/preemptible-vms/}.

\bibitem{propresp}
L.~Zhang, ``{Proportional response dynamics in the Fisher market},''
  \emph{Theoretical Computer Science}, vol. 412, no.~24, pp. 2691--2698, May
  2011.

\bibitem{She95}
S.~Shenker, ``{Fundamental Design Issues for the Future Internet},'' \emph{IEEE
  Journal of Selected Areas in Communications}, vol.~13, no.~7, pp. 1176--1188,
  Sep. 1995.

\bibitem{HZC07}
P.~Hande, S.~Zhang, and M.~Chiang, ``{Distributed rate allocation for inelastic
  flows},'' \emph{IEEE/ACM Transactions on Networking}, vol.~15, no.~6, pp.
  1240--1253, Dec. 2007.

\bibitem{Mo00}
J.~Mo and J.~Walrand, ``{Fair end-to-end window-based congestion control},''
  \emph{IEEE/ACM Trans. Networking}, vol.~8, no.~5, pp. 556--567, Oct. 2000.

\bibitem{mora}
P.~Caballero and et. al., ``{Multi-Tenant Radio Access Network Slicing:
  Statistical Multiplexing of Spatial Loads},'' \emph{IEEE/ACM Trans.
  Networking}, vol.~25, no.~5, pp. 3044--3058, Oct. 2017.

\bibitem{BeT89}
D.~Bertsekas and J.~Tsitsiklis, \emph{Parallel and distributed computation:
  numerical methods}.\hskip 1em plus 0.5em minus 0.4em\relax Prentice hall
  Englewood Cliffs, NJ, 1989.

\bibitem{IMT09}
{ ITU-R}, ``{Report ITU-R M.2135-1, Guidelines for evaluation of radio
  interface technologies for IMT-Advanced},'' Technical Report, Dec 2009.

\bibitem{Ye13}
Q.~Ye and et. al., ``{User Association for Load Balancing in Heterogeneous
  Cellular Networks},'' \emph{IEEE Trans. Wireless Comm.}, vol.~12, no.~6, pp.
  2706--2716, Jun. 2013.

\bibitem{TS36213}
{3GPP}, ``{Evolved Universal Terrestrial Radio Access (E-UTRA); Physical layer
  procedures},'' TS 36.213, v12.5.0, Rel. 12, Mar. 2015.

\bibitem{HLV06}
E.~Hyytia, P.~Lassila, and J.~Virtamo, ``{Spatial node distribution of the
  random waypoint mobility model with applications},'' \emph{IEEE Trans. Mobile
  Computing}, vol.~5, no.~6, pp. 680--694, June 2006.

\bibitem{slaw}
K.~Lee and et. al., ``{SLAW: self-similar least-action human walk},''
  \emph{IEEE/ACM Trans. Networking}, vol.~20, no.~2, pp. 515--529, Apr. 2012.

\bibitem{gps}
A.~K. Parekh and R.~G. Gallager, ``{A generalized processor sharing approach to
  flow control in integrated services networks: The single-node case},''
  \emph{IEEE/ACM Trans. Networking}, vol.~1, no.~3, pp. 344--357, Jun. 2004.

\bibitem{convexopt}
S.~Boyd and L.~Vandenberghe, \emph{Convex Optimization}.\hskip 1em plus 0.5em
  minus 0.4em\relax Cambridge University Press, 20042.

\end{thebibliography}

\end{document}